\documentclass[aps,prl,twocolumn,groupedaddress]{revtex4}
\usepackage{graphicx}
\usepackage{amsfonts,amssymb,latexsym}
\begin{document}

\title{Relationship between non-exponentiality of relaxation and
relaxation time at the glass transition}
\author{K. Trachenko,$^{1}$, C. M. Roland$^{2}$ and R.
Casalini,$^{2,3}$\\
$^{1}$Department of Earth Sciences, University of Cambridge, UK\\
$^{2}$Naval Research Laboratory, Washington, DC 20375-5342, USA\\
$^{3}$George Mason University, Fairfax, VA 22030, USA\\}

\begin{abstract}
By analyzing the experimental data for various glass-forming liquids and polymers, we find that non-exponentiality $\beta$ and
the relaxation time $\tau$ are uniquely related: $\log(\tau)$ is an approximately linear function of $1/\beta$, followed by a
crossover to a higher linear slope. We rationalize the observed relationship using a recently developed approach, in which the
problem of the glass transition is discussed as the elasticity problem.
\end{abstract}

\maketitle

Freezing a liquid to obtain glass may seem a familiar and conceptually simple process, and yet its theoretical description
remains elusive. Such a description, as widely perceived, should provide a consistent theory for the two main properties
that a liquid acquires in the glass transformation range: non-exponential relaxation and super-Arrhenius temperature
dependence of relaxation time \cite{dyre}.

At high temperature, a liquid under external perturbation relaxes exponentially fast: a relaxing quantity $q(t)$ decays as
$\exp(-(t/\tau))$, where $\tau$ is associated with the transition over a single activation barrier. This is known as Debye
relaxation. On lowering the temperature, relaxation changes markedly, and is described by a stretched-exponential function,
$q(t)\propto \exp(-(t/\tau)^\beta)$, where $0<\beta<1$ \cite{dyre,phillips}. The transition from Debye relaxation to
stretched-exponential relaxation (SER) marks the onset of glass transformation range. The transformation is complete when,
by convention, the relaxation time $\tau$ increases to the experimental time scale of 100--1000 seconds, corresponding to
the glass transition temperature $T_g$. In the glass transformation range, $\tau$ often increases faster than Arrhenius, and
is well approximated by the Vogel-Fulcher-Tamman (VFT) law, $\tau=\tau_0\exp(A/(T-T_0))$ \cite{dyre}.

$\beta$ and $\tau$ are therefore two fundamental parameters that describe a liquid in the glass transformation range. A
challenge for a theory of the glass transition is to propose a description of these parameters. $\beta$ and $\tau$ have been
discussed in a number of popular theoretical approaches \cite{dyre,phillips,adam,volume,ander,domi}. An outstanding feature
of these approaches is that $\beta$ and $\tau$ are often treated separately. One group of theories has offered the mechanism
of the increase of $\tau$, and includes the Adam-Gibbs entropy theory \cite{adam}, free volume theory \cite{volume}, elastic
models and other approaches (for a recent review, see Ref. \cite{dyre}). Another group of theories has derived $\beta$ for
SER (see, e.g., Refs. \cite{phillips,ander,domi}).

In view of this, it remains unclear what the relationship between $\beta$ and $\tau$ is, or if one exists at all. On the
other hand, because the increase of non-exponentiality and relaxation time are the two signatures of glass transformation,
it is natural to ask if there exists a fundamental process of slowing down of molecular motion that affects both quantities.
If a single mechanism affects both $\beta$ and $\tau$, it should be reflected in a well-defined relationship between these
parameters at different temperatures in the glass transformation range.

In this paper, we show that a universal relationship between $\beta$ and $\tau$ exists for all temperatures in the glass
transformation range: $\log(\tau)$ is an approximately linear function of $1/\beta$, followed by a crossover to a higher linear
slope. We discuss the observed behaviour in the elastic picture of the glass transition.

We have analyzed the experimental data on dielectric relaxation, including our recent results, as well as earlier data
\cite{sti,sti1,bloc,casa1,dixon1,casa2,sekula,roland,roland1,berber,schu,qi,roland2,paluch,casa3,casa}. At each temperature,
$\beta$ and $\tau$ were determined from the location and width of the dielectric loss peak, respectively. Prompted by our recent
work on glass transition \cite{beta,tau}, we have plotted $\log(\tau)$ as a function of $1/\beta$. In Figure 1, this dependence
is shown for 15 different glass-forming systems.

\begin{figure}
{\scalebox{0.9}{\includegraphics{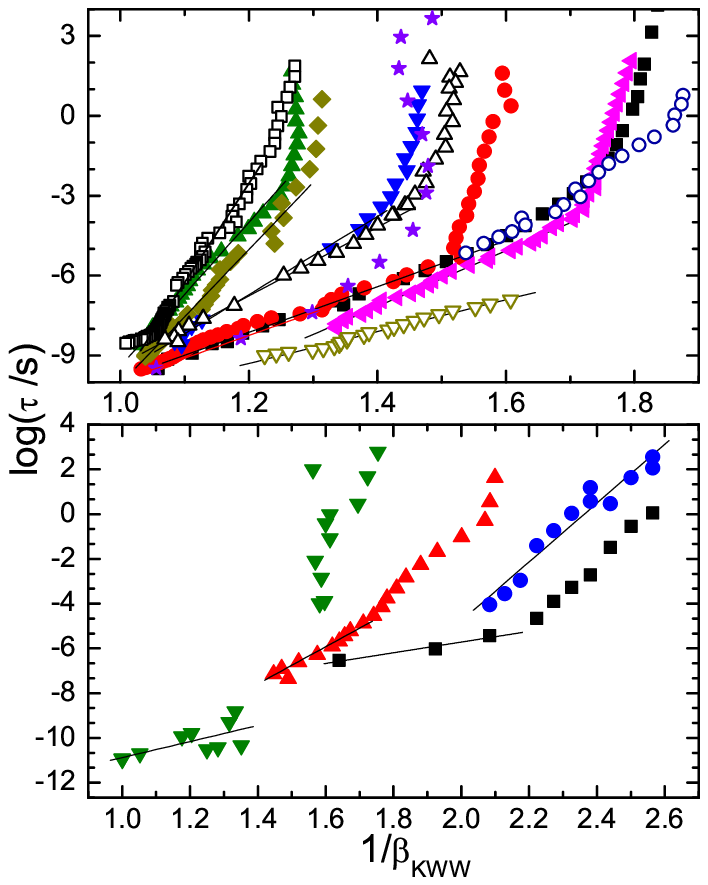}}} \caption{$\log(\tau)$ vs $1/\beta$ for 15 glass-forming systems. Upper
panel: molecular liquids ($\blacksquare$ o-terphenyl \cite{sti}, $\bullet$ salol \cite{sti1}, $\blacktriangle$ PDE
\cite{sti1}, $\bigstar$ 4-tertbutyl pyridine \cite{bloc}, $\blacktriangledown$ 54 \% chlorinated biphenyl \cite{casa1},  ,
$\blacklozenge$ propylene carbonate \cite{sti1}, $\blacktriangleleft$ dibutylphthalate \cite{dixon1}, $\square$ KDE
\cite{sti},  $\circ$ BMMPC \cite{sti}, $\bigtriangleup$ 62\% chlorinate biphenyl \cite{casa2}, $\bigtriangledown$
diisobutylphthalate \cite{sekula}). Lower panel: $\blacktriangle$ polyvinylacetate \cite{sti}, $\bullet$
polycyclohexylmethacrylate \cite{roland}, $\blacksquare$ polyvinylethylene \cite{roland1}, $\blacktriangledown$ bromopentane
\cite{berber}. Lines are guides for the eye. These represent the materials for which data are available covering a broad
range and exhibiting ``normal'' behavior (i.e., $\beta$ decreases with decreasing temperature). There are some glass-formers
(e.g., NMEC \cite{schu}, MTHF \cite{qi}, 1,4-polyisoprene \cite{roland2}, polymethylphensiloxane \cite{paluch}, 1-propanol
\cite{sti1}, ethanol \cite{sti1}) for which $\beta$ is almost constant over the supercooled regime, and correspondingly
there is a weak or absent crossover. In other materials, a neighboring dispersion, such as the normal mode (e.g.,
polyoxybutylene \cite{casa3}) or a secondary peak (BMPC \cite{sti})), obfuscates accurate analysis of the $\alpha$-process.
For propylene glycol and glycerol \cite{dixon1}, $\log(\tau)$ changes with $\beta$, but does not show the linear behaviour
as in this Figure.}
\end{figure}

Figure 1 immediately highlights our central point, namely that a universal relationship $f(\beta,\tau)=0$ can be identified for
the studied systems in the entire range of glass transformation. First, at high temperature, $\log(\tau)$ is approximately
proportional to $1/\beta$. Second, as the temperature is reduced, a crossover to another higher slope takes place.

We note here that generally, presenting the data as in Figure 1 is attractive since it does not require scaling by $T_g$, an
arbitrary quantity from a theoretical point of view. In addition to providing a universal relationship between $\beta$ and
$\tau$, the plot in Figure 1 can serve as a convenient comparative representation of non-exponentiality in different systems
approaching their glass transition.

We also note that the universality of the pattern in Figure 1 is extended to the domain of high pressure. In the dielectric
spectroscopy experiment, the increase of $\tau$ at higher pressure can be counter-balanced by an increase of temperature. For
many systems, it has been shown \cite{ngai1} that different combinations of pressure and temperature that keep $\tau$ constant,
always give the same value of $\beta$. This signifies the universal relationship between $\beta$ and $\tau$.

There is no {\it a-priori} reason why the location of the dielectric loss peak ($\tau$) should be correlated with its width
($\beta$). Hence the existence of a universal relationship between $\beta$ and $\tau$ strongly suggests that they reflect
the same slowing-down mechanism operative in the glass transformation range.

Figure 1 presents a challenge for a theory of the glass transition. In order to rationalize the observed behaviour, one requires
an approach to the glass transition that offers descriptions of both $\beta$ and $\tau$. Some have been proposed
\cite{lubch,garrah}, and it would be interesting to see the predictions of these approaches regarding the observed behaviour.
Here, we discuss the observed relationship based on the recently developed approach, in which the problem of the glass
transition is discussed as the problem of elasticity \cite{tau,beta}. Below we briefly review the model, followed by a
discussion of its prediction about the relationship between $\beta$ and $\tau$.

A glass is different from a liquid by virtue of its ability to support shear stresses on experimental time scales \cite{dyre}.
In our picture, we discuss that when considering stress relaxation, a liquid can be treated as an elastic medium. In other
words, we show that the problem of glass transition can be formulated as the elasticity problem. Important to this discussion is
the elastic feed-forward interaction between local relaxation events (LREs) \cite{beta,tau}. Lets consider the dynamics of LREs,
induced in a liquid by an external perturbation (e.g. shear stress). Some time ago, Orowan's introduced terms of ``concordant''
and ``discordant'' events \cite{orowan}: a concordant local rearrangement is accompanied by a strain agreeing in direction with
the applied external stress, and reduces the energy and local stress. A discordant rearrangement, on the other hand, increases
the energy and local stress. This has led to a general result that stress relaxation by earlier concordant events leads to the
increase of stress on later relaxing regions in a system. Goldstein applied the same argument to a viscous liquid \cite{gold}:
consider a system under external stress which is counterbalanced by stresses supported by local regions. When a local
rearrangement to a potential minimum, biased by the external stress, occurs (a concordant event), this local region supports
less stress after the event than before; therefore, other local regions in the system should support more stress after that
event than before \cite{gold}.

Let $\Delta p$ be the increase of shear stress on a current LRE due to previous concordant LREs. If $n$ is the current number of
LREs, $\Delta p$ is a monotonically increasing function of $n$. The increase of stress, on a currently relaxing region increases
its activation barrier $V$. It has been argued that $V$ is given by the elastic shear energy of a surrounding liquid
\cite{dyre1}. This result was confirmed by the experimental measurements of the shear modulus, showing that the activation
barrier increases with the shear energy \cite{dyre1}. Because, as discussed by Orowan and Goldstein, previous LREs reduce stress
in the direction ``concordant'' to the external stress, the increase of shear stress on later rearranging regions consistently
increases shear strain on them in the same direction, increasing shear energy and therefore $V$. The increase of $V$ due to the
additional stress $\Delta p$, $\Delta V$, is given by work $\int \Delta p {\rm d}q$. If $q_a$ is the characteristic volume
\cite{dyre1}, $\Delta V=\Delta p q_a$, and we find $V=V_0+q_a\Delta p$, where $V_0$ is the high-temperature activation barrier.

Because $\Delta p$ is a monotonically increasing function of $n$ and $V=V_0+q_a\Delta p$, we find that $V$ is also a
monotonically increasing function of $n$. This provides the {\it feed-forward interaction mechanism} between LREs, in that
activation barriers increase for later events.

It is important to discuss how the feed-forward interaction mechanism operates on lowering the temperature. The elastic wave
that propagates stress created by a LRE is stopped when another LRE takes place at the front of the wave. If $\tau$ is the
structural relaxation time, the elastic wave propagates without dissipation distance $d=c\tau$, where $c$ is the speed of sound.
$d$ can therefore be called the liquid elasticity length. Because $c$ is on the order of $a/\tau_0$, where $a$ is the
interatomic separation of about 1 \AA\ and $\tau_0$ the oscillation period, or inverse of Debye frequency ($\tau_0=0.1$ ps),

\begin{equation}
d=a\frac{\tau}{\tau_0}
\end{equation}

Let $d_m$ be the distance between neighbouring LREs. $d_m$ is the distance between the centres of neighbouring molecular cages
of about 10 \AA. At high temperature, when $\tau\approx\tau_0$, $d<d_m$ (see Eq. (1)). This means that neighbouring LREs do not
elastically interact. Because events are independent, we obtain the expected high-temperature result that relaxation is
exponential in time and Arrhenius in temperature. Because $\tau=\tau_0\exp(V/kT)$, a certain temperature always gives the
opposite condition, $d>d_m$, at which point the elastic feed-forward interaction mechanism between LREs becomes operative. Note
that the maximal time between two neighbouring LREs is given by $\tau$, hence $d=c\tau<d_m$ ($d=c\tau>d_m$) also means that the
time between the neighbouring LREs is shorter (longer) than the time of elastic propagation between the events $d_m/c$. This is
another way of showing that local events relax as independent at high temperature, but start to interact as the temperature is
lowered. We have recently shown that this results in the crossover from exponential to stretched-exponential relaxation
\cite{beta}. The crossover temperature $T_c$ is calculated by putting $d=d_m$ in Eq. (1).

To calculate how $V$ depends on the current number of LREs $n$, we introduce the dynamic variable $n(t)$, the current number of
relaxing events induced by an external perturbation (i.e., in addition to thermally-induced events). We consider relaxation at
constant temperature. According to Eq. (1), this sets the range of the feed-forward interaction $d$. $n(t)$ starts from zero and
increases to its final value $n_{\rm r}$, $n(t)\rightarrow n_{\rm r}$ as $t\rightarrow\infty$. Lets consider the current LRE to
relax in the centre of the sphere of diameter $d$. As discussed above, all previous remote concordant LREs within distance $d$
from the centre participate in the feed-forward interaction, increasing stress $\Delta p$ on the central region and hence
increasing $V$ for the central LRE. $\Delta p$ can be calculated by integrating the contributions of remote concordant LREs. A
straightforward integration, together with $V(n)=V_0+q_a\Delta p$ from the above, gives \cite{beta}:

\begin{equation}
V(n)=V_0+V_1\frac{n}{n_r}
\end{equation}
\noindent where $V_1=\pi/2\rho_r q_a\Delta p_0 d_0^3\ln(2d/d_0)$, $d_0$ is on the order of the size of a relaxing region,
$\rho_r$ is the density of relaxing regions, $\rho_r=6n_r/\pi d^3$, and $\Delta p_0$ is the decrease of stress due to a remote
concordant LRE.

In Eq. (2), $V_1$ depends on temperature through $d$ (see Eq. (1)). As we have recently shown \cite{tau}, using
$\tau=\tau_0\exp(V/kT)$ in Eq. (1) and eliminating $d$ from Eq. (2), gives the VFT law for $\tau$. Here, $\tau$ corresponds to
$n=n_r$, i.e. maximal relaxation time of the system \cite{tau}.

Eq. (2) describes the feed-forward interaction mechanism in a liquid at $T<T_c$. We are now set to write the equation that
relates $\beta$ and $\tau$. The rate of LREs, ${\rm d}n/{\rm d}t$, is proportional to the number of unrelaxed events, $(n_{\rm
r}-n)$, and the event probability, $\rho=\exp(V/kT)$. Since $V$ depends on $n$ (see Eq. (2)), $\rho$ becomes dependent on $n$.
Introducing $q=n/n_r$, and reduced time $t/\tau_0$, we write:

\begin{equation}
\frac{{\rm d}q}{{\rm d}t}=(1-q)\exp\left(-\frac{V_0+V_1q}{kT}\right)
\end{equation}

Eq. (3) has two parameters, $V_0/kT$ and $\alpha=V_1/kT$. We have recently shown \cite{beta} that its solution is well
approximated by the two-parameter SER, $q(t)=1-\exp{(-(t/\tau)^\beta})$. Whereas $\tau$ depends on both $V_0/kT$ and $\alpha$,
$\beta$ depends on $\alpha$ only; the smaller $\alpha$, the larger $\beta$ ($\alpha=0$ gives exponential relaxation, $\beta=1$).
We solve Eq. (3) for different values of $\alpha$, fit the solution to the form of SER above, and find that $1/\beta=1+C\alpha$,
where $C$ is a constant (see Figure 2). Joining this result with $V_1\propto\ln(d)\propto\ln(\tau)$ (see Eqs. 1-2), we find that
$\ln(\tau)$ is a quasi-linear function of $1/\beta$.

\begin{figure}
\rotatebox{-90}{\scalebox{0.5}{\includegraphics{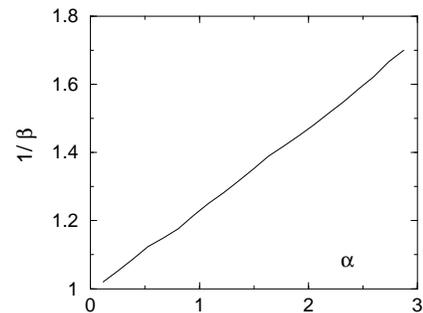}}} \caption{$\beta$ as a function of $\alpha$, obtained from the
solution of Eq. (3).}
\end{figure}

To discuss the crossover to the second higher slope in Figure 1, we note that our picture predicts crossovers of both $\tau$ and
$\beta$ when at low temperature $d$ reaches the system size $L$. The crossover of $\tau$ takes place because $V$ is proportional
to $V1\propto\ln(d)$ (see Eq. (2)) \cite{tau}. As long as $d<L$, $V$ increases with temperature, because $d$ is
temperature-dependent (see Eq. (1)), giving the VFT dependence of relaxation time \cite{tau}. When $d\ge L$, $V$ can not
increase by way of increasing $d$, resulting in a weak temperature dependence of $V$ at $d\ge L$ and thus the crossovers
observed experimentally \cite{schon,sti1}. The same reasoning is applied to the crossover of $\beta$. $V_1$ increases on
lowering the temperature because $V_1\propto \ln(d)$ in Eq. (2). This remains true as long as, on lowering the temperature,
$d<L$. When $d\ge L$, $V_1\propto\ln(L/d_0)$, and is temperature-independent. Hence at $d=L$, $V_1$ shows a kink and starts to
saturate to a constant value. Because $\beta$ decreases with $V_1/kT$ (see Figure 2), we find that $d=L$ should mark the
crossover of $\beta$ to the lower slope. Using salol as an example, $d=L\approx 1$ mm in Eq. (1), and the VFT parameters $A=839$
K and $T_0=195$ K for $\tau$ \cite{casa}, we find the crossover temperature of 247 K. This agrees well with the crossover of
$\beta$ shown in Figure 3.

Because our picture predicts the crossover of $\beta$ to the lower slope, the crossover of $\log(\tau)$ vs $1/\beta$ to the
higher slope also takes place, as is seen in Figure 1.

The crossover of $\beta$ at $d=L$ is expected to be more pronounced than that of $\tau$, because the activation barrier is the
sum of a constant term and a temperature-dependent term $\ln(d/d_0)$ \cite{tau}, whereas $\beta$ is solely defined by
$\alpha=V_1/kT$ (see Eqs. (2-3)). Consistent with this, we find that the crossover in Figure 1 is affected mostly by the
crossover of $\beta$. In Figure 3 we observe that for salol at low temperature, $1/\beta$ trails off at about 1.5, the same
value of $1/\beta$ associated with the crossover to the higher slope in Figure 1. The crossover in Figure 1 takes place in the
temperature range in which $\tau$ shows a crossover as well, although the latter is less pronounced as compared with the
crossover of $\beta$.

\begin{figure}
{\scalebox{0.9}{\includegraphics{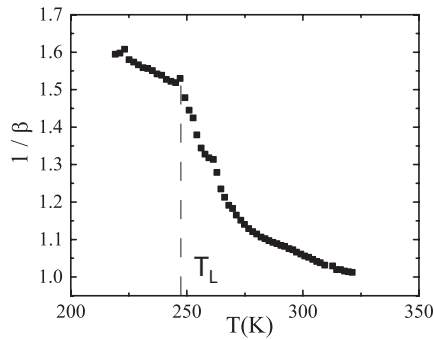}}} \caption{Temperature dependence of $1/\beta$ in salol. $T_L$ is the
temperature of the crossover when $d=L$.}
\end{figure}

In summary, we have shown that the non-exponentiality parameter $\beta$ and relaxation time $\tau$ are uniquely related in the
glass transformation range, and rationalized this relationship in the approach that discusses the problem glass transition from
the elastic point of view.

We are grateful to Prof. R. B\"{o}hmer and E. R\"{o}essler for providing their data in digital form, and to EPSRC and the
Office of Naval Research for support.

\bibliography{scibib}

\begin{thebibliography}{99}

\bibitem{dyre} J. C. Dyre, Rev. Mod. Phys. {\bf 78}, 953 (2006).

\bibitem{phillips} J. C. Phillips, Rep. Prog. Phys. {\bf 59}, 1133
(1996).

\bibitem{adam} G. Adam and J. H. Gibbs, J. Chem. Phys. {\bf 43}, 139 (1965).

\bibitem{volume} M. H. Cohen and D. Turnbull, J. Chem. Phys. {\bf 31}, 1164 (1959).

\bibitem{ander} R. G. Palmer, D. L. Stein, E. Abrahams and P. W. Anderson, Phys.
Rev. Lett. {\bf 53}, 958 (1984).

\bibitem{domi} C. D. Dominics, H. Orland, and F. Lainee, J. Phys. (France) Lett. {\bf 46}, L463 (1985).

\bibitem{sti} F. Stickel, Ph.D. thesis, Mainz University, Germany (Shaker, Aachen, 1995).

\bibitem{sti1} F. Stickel, E. W. Fischer and R. Richert, J. Chem. Phys. {\bf 104}, 2043 (1996).

\bibitem{bloc} T. Blochowicz et al, J. Chem. Phys. {\bf 124}, 134503 (2006).

\bibitem{casa1} R. Casalini, P. G. Santangelo and C. M. Roland, J. Phys. Chem. B {\bf 106}, 11492 (2002).

\bibitem{dixon1} P. K. Dixon, L. Wu, S. R. Nagel, B. D. Williams and J. P. Carini, Phys. Rev. Lett.
{\bf 65}, 1108 (1990).

\bibitem{casa2} R. Casalini, M. Paluch, J. J. Fontanella and C. M. Roland, J. Chem. Phys. {\bf 117}, 4901 (2002).

\bibitem{sekula} M. Sekula et al, J. Phys. Chem. B {\bf 108}, 4997 (2004).

\bibitem{roland} C. M. Roland and R. Casalini, Macromolecules, in press.

\bibitem{roland1} C. M. Roland et al, Macromolecules {\bf 36}, 4954 (2003).

\bibitem{berber} J. G. Berberian and R. H. Cole, J. Chem. Phys. {\bf 84}, 6921 (1986).

\bibitem{schu} J. Sch\"{u}ller, R. Richert and E. W. Fischer, Phys. Rev. B {\bf 52}, 15232 (1995).

\bibitem{qi} F. Qi et al, J. Chem. Phys. {\bf 118}, 7431 (2003).

\bibitem{roland2} C. M. Roland, M. J. Schroeder, J. J. Fontanella and K. L. Ngai, Macromolecules {\bf 37}, 2630 (2004).

\bibitem{paluch} M. Paluch, C. M. Roland and S. Pawlus, J. Chem. Phys. {\bf 116}, 10932 (2002).

\bibitem{casa3} R. Casalini and C. M. Roland, Macromolecules {\bf 38}, 1779 (2005).

\bibitem{casa} R. Casalini, K. L. Ngai and C. M. Roland, Phys. Rev. B {\bf 68}, 014201 (2003).

\bibitem{beta} K. Trachenko, Phys. Rev. B {\bf 75}, 212201 (2007).

\bibitem{tau} K. Trachenko, cond-mat/0704.2975v1.

\bibitem{ngai1} K. L. Ngai, R. Casalini, S. Capaccioli, M. Paluch and C. M. Roland, Journal of Physical Chemistry B {\bf 109},
17356 (2005).

\bibitem{lubch} V. Lubchenko and P. G. Wolynes, J. Chem. Phys. {\bf 121}, 2852 (2004).

\bibitem{garrah} J. P. Garrahan and D. Chandler, Proc. Nat. Acad. Sci. USA {\bf 100}, 9710 (2003).

\bibitem{orowan} E. Orowan, Proceedings of the First National Congress
of Applied Mechanics (American Society of Mechanical Engineers, New York), 453 (1952).

\bibitem{gold} M. Goldstein, J. Chem. Phys. {\bf 51}, 3728 (1969).

\bibitem{dyre1} J. C. Dyre, N. B. Olsen and T. Christensen, Phys. Rev. B {\bf 53}, 2171 (1996).

\bibitem{schon} A. Sch\"{o}nhals, Europhys. Lett. {\bf 56}, 815 (2001).

\end{thebibliography}

\bibliographystyle{Science}

\end{document}